\documentstyle[preprint,aps]{revtex}
\begin{document}

\begin{title}
{
Spin-Charge Separation is the Key to the High $T_c$ cuprates
}
\end{title}

\author{ Philip W. Anderson}

\address {Joseph Henry Laboratories of Physics\\
Princeton University, Princeton, NJ 08544}
\maketitle


\vfill\eject

The most striking fact about the high-$T_c$ cuprates is that in none
of the relevant regions of the phase diagram is there any
evidence of the usual effects of phonon or impurity scattering.
This is strong evidence that these states are in a ``quantum
protectorate'', to borrow Laughlin's term. (Two striking
experimental facts which demonstrate this are the absence of
phonon self-energy in ARPES measurements, demonstrated recently by
Johnson in BISCO, and the scattering-independent $T_c$ in YBCO, even
though the superconductivity is $d$-wave.) A quantum protectorate
is a state in which the many-body correlations are so strong that
the dynamics can no longer be described in terms of individual
particles, and therefore perturbations which scatter individual particles
are not effective.

The Mott-Hubbard antiferromagnetic phase is manifestly
spin-charge separated (there is a charge gap, but no spin gap),
and I propose this property extends throughout the phase diagram
in different guises, and is the reason for the quantum
protectorate. Quasiparticles are never the exact, long-lived
elementary excitations ($Z\equiv 0$ throughout) so that
scattering of electrons does not necessarily disturb the
excitations, especially the spinons, the  Fermion-like
elementary magnetic excitations with spin 1/2 and charge 0.

I want to emphasize that this protectorate effect is completely
incompatible with any perturbative theory starting from a Fermi
liquid approach, as for example the spin-fluctuation theory. The
experimental situation presents us with a clean dichotomy, which
cannot be repaired by ``summing all the diagrams''.

I propose that as we dope the antiferromagnetic state, two things
happen. First, there is a first-order phase transition in the
charge sector where the Mott-Hubbard gap closes, which sometimes leads to
mesoscopic inhomogeneity (the stripe phenomenon). Second, and
independently, we pass a critical point in the spin sector where
antiferromagnetism vanishes and becomes a soft mode of the
$d$-wave (or flux-phase-like) RVB.

T. Hsu, in his thesis, long ago showed that antiferromagnetism
could be treated as an unstable mode of the ``flux phase'' i.e.,
the $d$-wave RVB. Hence when the latter becomes stable
antiferromagnetism becomes a stable soft mode, which is seen as
the notorious neutron resonance of Keimer.

This RVB, which was postulated independently by
Affleck-Marston-Kotliar and by Laughlin, is the spin gap phase,
stable below a crossover $T^*$ which is a rapidly decreasing
function of doping. It is correct to think of this phase as analogous to
the Mott insulator: where the Mott phase has a charge gap and no spin gap,
this one has a spin gap--for most momenta--and no charge gap.

What is the phase $\underline{\rm above}$ $T^*$? This is not a conventional
Fermi
liquid, but the original ``extended $s$-wave'' RVB I proposed in
1987, equivalent to the ``tomographic Luttinger liquid'' I
derived for moderate densities in 1989. If there is an
antiferromagnetic superexchange $J$ this phase has a Cooper
instability at $T^*$ where the $d$-wave spin gap is favored.

Why does $T^*$ decrease so rapidly with doping? The Cooper
instability occurs at
$$kT^* \sim p_Fv_s \exp {(-p_Fv_s)\over J_{eff}}$$
where $v_s$ is the spinon velocity. When $x\to 1$, the spinon
and holon velocities become equal, and equal to $t/p_F$, while as
$x\to 0$, $v_s=J/p_F$, so we linearly interpolate
$$p_Fv_s\simeq J+x(t-J).$$
As for the interaction term, the antiferromagnetic interaction
will tend to be compensated by a ferromagnetic double exchange
term roughly proportional to $x$, which comes from loop
(non-repeating) paths of the holes. For rough purposes, I
estimate
$$J_{eff}(x)=J-tx.$$
This is adequate to account for the vanishing of $T^*$ at around
$x\simeq .3$. This Cooper instability is not a true phase
transition. Both the $s$-wave state and the $s+id$ retain the
full symmetry of the Hamiltonian,
$$[SU(2)]_{spin} \times [U(1)]_{charge}\times [\rm Lattice\
Symmetry].$$
It is the conventional Fermi liquid which has an anomalous extra
symmetry $Z_2$ mixing spin and charge: the Fermi liquid is itself
a quantum critical point.

The RVB (spin gap) phase is not superconducting.  The motivation which
drives superconductivity and converts the spin gap into a
superconducting gap is frustrated kinetic energy.  The opening up
of a gap in the spectrum at the Fermi level means that the
distribution $n(k)$ cannot approach its optimal step-function
form (or step-function-like, in the case of a Luttinger liquid).
This effect may be quantified by using the Ferrell-Glover-Tinkham
sum rule.

$$\int^\infty_0\sigma(\omega)d\omega=e^2<K.E.>$$
It was observed by Orenstein et al, very early, that in the
underdoped (spin-gapped) cuprates a gap in the optical
conductivity opens up. The magnitude of
$\int \sigma(\omega)d\omega$ is proportional to
$x$ (as noted by Sawatsky) so the
relevant loss of kinetic energy is $\propto x$. (P.A. Lee has
estimated a similar effect quantitatively.) One may think of this
effect as setting an upper limit to the transition temperature
$T_C <T_{KE}\propto x.$ But if the spin gap $T^*$ is smaller than
this upper limit, essentially no spin gap can open without a
charge gap as well, so $T_c$ follows $T^*$ down at high doping.

The above is far from a quantitative theory of $T_c$. In
particular, I believe there is still a role for the interlayer
kinetic energy in the bilayer and multilayer cases. But contrary
to my ``Dogma V'' there is a one-layer mechanism for
superconductivity which seems to be quite effective in some cases.
But in all cases the relevant mechanism is the recovery of
frustrated kinetic energy.  Note that in the charge channel this is not a
d-wave but an s-wave condensation, hence not strongly affected by ordinary
scattering.

To summarize:  The two-dimensional electron gas in the cuprates is
dominated by the short-range repulsive interaction which remains relevant
and causes spin-charge separation.  A spin gap develops in the metallic
phase below a crossover temperature T*, at the  Cooper instability caused
by the antiferromagnetic superexchange. The extra kinetic energy required
to open the spin gap is relaxed at a lower temperature Tc by making the
charge fluctuations coherent, and this is the immediate cause of
superconductivity.

\end{document}